\documentclass{appolb}
\usepackage{cite}
\usepackage{hyperref}
\usepackage{graphicx}
\usepackage{amsmath}
\usepackage[dvipsnames]{xcolor}
\usepackage{animate}
\usepackage{authblk}
\newtheorem{theorem}{Theorem}[section]
\DeclareMathOperator{\sgn}{sgn}

\begin{document}

\title{Timelike and null geodesics in the Schwarzschild space-time: Analytical solutions%
\thanks{Talk presented at the 8th Conference of the Polish Society on Relativity (POTOR8), 19-23 September 2022, Banach Center IMPAN, Warsaw, Poland \cite{prezentacja}.}%
}
\author{Adam Cie\'{s}lik$^{1,2}$, Patryk Mach$^1$
\address{
     $^1$Instytut Fizyki Teoretycznej, Wydział Fizyki, Astronomii i Informatyki Stosowanej, Uniwersytet Jagiello\'{n}ski, {\L}ojasiewicza 11, 30-348 Kraków, Poland \\
     $^2$Szkoła Doktorska Nauk Ścisłych i Przyrodniczych, Uniwersytet Jagiello\'{n}ski
    }
}

\maketitle
\maketitle
\begin{abstract}
The theory of Schwarzschild geodesics is revisited. Using a theorem due to Weierstrass and Biermann, we derive concise formulas describing all timelike and
null trajectories in terms of Weierstrass elliptic functions. The formulation given in this note uses an analogue of the so-called Mino time.
\end{abstract}

\section{Introduction and motivation}

Since the discovery of the Schwarzschild metric, different ways of expressing solutions to Schwarzschild geodesic equations have been developed \cite{hagihara_theory_1930, mielnik_study_1962, chandrasekhar_mathematical_1983, scharf_schwarzschild_2011, kostic_analytical_2012, CM2022}. In most cases, they differ in the assumed parametrization and in the types of elliptic functions used to write the solution.

\tolerance=5000
Our motivation for revisiting the theory of Schwarzschild geodesics is related with works on the kinetic description of the Vlasov gas in the Schwarzschild spacetime and the accretion of the Vlasov gas onto Schwarzschild black holes \cite{rioseco_accretion_2017, rioseco_spherical_2017, mach_odrz_accretion_2021, mach_odrz_accretion_2021-1,mach_odrz_acta,Olivier_disks,gamboa_2021}, Reissner-Nordstr\"{o}m black holes \cite{cieslik_2020} and Kerr black holes \cite{Olivier3, Andersson, CMO2022, mach_odrz_acta_2023}. In particular, aiming at constructing Monte Carlo type simulations of the gas consisting of non colliding particles moving around the Schwarzschild black hole, we were searching for a concise description of all types of time-like and null trajectories in the Schwarzschild metric, which would depend explicitly on standard constants of motion (in particular, the energy and the angular momentum) together with the initial location of the particle. It turns out that such a description can be written in terms of Weierstrass elliptic functions, basing on a theorem due to Weierstrass and Biermann \cite{Biermann_1865}. A detailed account of these results can be found in our recent paper \cite{CM2022}. This note is intended to give a short summary, but, in comparison to \cite{CM2022}, we use a different parametrization. In \cite{CM2022} solutions are expressed in terms of the so-called true anomaly; in this note we use an analogue of the Mino time, introduced originally for Kerr geodesics in \cite{mino_perturbative_2003}.

\section{Geodesic equations}
We work in standard Schwarzschild coordinates. The metric reads
\begin{equation}\label{metric}
g = -N d t^2 + \frac{d r^2}{N} + r^2 d \theta^2 +  r^2 \sin^2 \theta d \varphi^2,
\end{equation}
where
\begin{equation}
N = 1 - \frac{2M}{r},
\end{equation}
and $M$ is the Schwarzschild mass. Geodesic equations can be written in the form \cite{CM2022}
\begin{subequations}
\label{eqsofmotion1}
\begin{alignat}{1}
	\frac{dr}{d\tilde{s}} &= \epsilon_r \sqrt{E^2 - U_{l,m}(r)},\\
    \frac{d\theta}{d\tilde{s}} &= \frac{\epsilon_\theta}{r^{2}} \sqrt{ l^{2} -\frac{l_{z}^2}{\sin^2 \theta}},\\	
    \frac{d\varphi}{d\tilde{s}} &=  \frac{l_{z}}{r^{2}\sin^2 \theta}, \\
    \frac{dt}{d\tilde{s}} &=  \frac{E}{N}, \label{dtdstilde}
\end{alignat}
\end{subequations}
where
\begin{equation}
U_{l,m}(r) = \left(1 - \frac{2M}{r} \right) \left( \delta_m + \frac{l^2}{r^2} \right)
\end{equation}
is the radial effective potential and $E$, $l \ge 0$, $l_z$ denote constant values of the particle energy, the total angular momentum, and the azimuthal component of the angular momentum, respectively. The constant $\delta_m$ is defined as 
\begin{equation}
    \delta_m = \begin{cases} m^2 & \text{for timelike geodesics,} \\ 0 & \text{for null geodesics,} \end{cases}
\end{equation}
and the signs $\epsilon_\theta = \pm 1$ and $\epsilon_r = \pm 1$, correspond to the directions of motion.

There is a convenient way to partially decouple the above equations, which also works for the geodesic motion in the Kerr spacetime, and which for the latter case was introduced by Mino in \cite{mino_perturbative_2003}. The trick is to reparametrize geodesics so that
\begin{equation}
\frac{dx^\mu}{d \bar s} = r^2 \frac{dx^\mu}{d \tilde s} \; \; \text{or} \; \; \tilde s = \int_0^{\bar s} r^2 ds.
\end{equation}
We will refer to the parameter $\bar s$ as the Mino time. Additionally, to simplify further calculations, we will work in dimensionless rescaled variables, defined as in \cite{rioseco_accretion_2017}, i.e.,
\begin{alignat}{7}
    t = M \tau, &\; r = M \xi, &\; p_r =  m \pi_\xi, &\; p_\theta  =  M m \pi_\theta, &\; E  =  m \varepsilon, &\; l =  M m \lambda, &\; l_z = M m \lambda_z.
\end{alignat}
A new Mino time $s$ is defined by
\begin{equation}
    \bar{s} = \frac{s}{M m}.
\end{equation}

In terms of these dimensionles variables, geodesic equations (\ref{eqsofmotion1}) can be written as
\begin{subequations}
\label{eqsofmotion3}
\begin{eqnarray}
    \frac{d\xi}{ds} & = &  \epsilon_r \xi^2 \sqrt{\varepsilon^{2} - U_\lambda(\xi)}\label{xi_mot},\\
    \frac{d\theta}{ds} & = & \epsilon_\theta \sqrt{ \lambda^{2} - \frac{\lambda_{z}^2}{\sin^2 \theta}}\label{theta_mot},\\    
    \frac{d\varphi}{ds} & = & \frac{\lambda_{z}}{\sin^2 \theta}\label{phi_mot},\label{varphi_mot}\\ 
    \frac{d\tau}{ds} & = & \frac{\varepsilon \xi^2}{N(\xi)},\label{tau_mot}
\end{eqnarray}
\end{subequations}
where $N(\xi) = 1 - 2/\xi$. The dimensionless radial potential reads
\begin{equation}
\label{eff_pot}
U_\lambda(\xi) = \left( 1 - \frac{2}{\xi} \right)\left(  \delta + \frac{\lambda^2}{\xi^2}\right) = \delta - \frac{2}{\xi} \delta + \frac{\lambda^{2}}{\xi^{2}} - \frac{2\lambda^{2}}{\xi^{3}},
\end{equation}
where
\begin{equation}
    \delta = \begin{cases} 1 & \text{for time-like geodesics,} \\ 0 & \text{for null geodesics.} \end{cases}
\end{equation}

\section{Biermann–Weierstrass theorem}\label{Biermann-Weierstrass}
Solutions given in this work are based on the following result due to Biermann and Weierstrass. The original formulation of this theorem has been published in \cite{Biermann_1865}. More detailed proofs can be found in \cite{Greenhill_1892, Reynolds_1989, CM2022}.

\begin{theorem}
Let 
\begin{equation}
    f(x) = a_0 x^4 + 4 a_1 x^3 +6 a_2 x^2 + 4a_3 x + a_4,
\end{equation}
be a quartic polynomial. Denote the invariants of $f$ by $g_2$ and $g_3$, i.e.,
\begin{subequations}
\label{invariants_theorem}
\begin{eqnarray}
        g_2 & \equiv &  a_0 a_4 - 4a_1 a_3 + 3 a_2^2, \\
        g_3 & \equiv & a_0 a_2 a_4 + 2a_1 a_2 a_3 -a_2^3 -a_0 a_3^2 - a_1^2 a_4.
\end{eqnarray}
\end{subequations}
Let
\begin{equation}
\label{zthm}
     z(x) = \int^x_{x_0} \frac{dx^\prime}{\sqrt{f(x^\prime)}},
\end{equation}
where $x_0$ is any constant, not necessarily a zero of $f(x)$. Then 
\begin{equation}
\label{glowne}
     x = x_0 + \frac{- \sqrt{f(x_0)} \wp'(z) + \frac{1}{2} f'(x_0) \left[ \wp(z) - \frac{1}{24}f''(x_0) \right] + \frac{1}{24} f(x_0) f'''(x_0)}{2 \left[ \wp(z) - \frac{1}{24} f''(x_0) \right]^2 - \frac{1}{48} f(x_0) f^{(4)}(x_0)},
\end{equation}
 and
 \begin{subequations}
 \label{BW_wp}
 
 \begin{eqnarray}
 \wp(z) & = & \frac{\sqrt{f(x) f(x_0)} + f(x_0)}{2(x-x_0)^2} + \frac{f'(x_0)}{4(x-x_0)} + \frac{f''(x_0)}{24},\\
\wp'(z) & = & \textstyle  - \left[ \frac{f(x)}{(x-x_0)^3} - \frac{f'(x)}{4(x-x_0)^2} \right] \sqrt{f(x_0)} -  \left[ \frac{f(x_0)}{(x-x_0)^3} + \frac{f'(x_0)}{4(x-x_0)^2} \right]\sqrt{f(x)}  \nonumber,\\
\end{eqnarray}
\end{subequations}
where $\wp(z) =\wp(z;g_2,g_3)$ is the Weierstrass function corresponding to invariants (\ref{invariants_theorem}).
\end{theorem}

\section{Solutions}
\label{sec:solution}

\subsection{Solution for \texorpdfstring{$\xi(s)$}{E(s)}}

Equation \eqref{xi_mot} can be rewritten in the Weierstrass form
\begin{equation}
\label{equationf}
   \frac{d\xi}{ds} =\epsilon_r \sqrt{f(\xi)},
\end{equation}
where
\begin{equation}
\label{f_general}
    f(\xi) = a_0 \xi^4 + 4a_1 \xi^3 +6a_2\xi^2 + 4a_3\xi + a_4,
\end{equation}
and
\begin{equation}
\label{fcoeffs}
    a_0 =\varepsilon^2 -\delta, \quad 4a_1 = 2 \delta, \quad 6a_2 = - \lambda^2, \quad 4a_3 =  2 \lambda^2, \quad a_4 = 0.
\end{equation}
For a segment of the trajectory from $\xi_0$ to $\xi$, we get
\begin{equation}
\label{s_integral}
s = \epsilon_r \int_{\xi_0}^\xi \frac{d \xi^\prime}{\sqrt{f(\xi^\prime)}},
\end{equation}
where $\xi_0$ is a reference radius corresponding to $s = 0$. Weierstrass invariants of the polynomial $f$ read 
\begin{subequations}
\label{invariants_phys}
\begin{eqnarray}   
        g_2 & = & \frac{1}{12}\lambda^4 - \delta  \lambda^2,   \\
        g_3 & = & \frac{1}{6^3} \lambda^6 - \frac{ \delta}{12}\lambda^4 - \frac{1}{4} \left( \varepsilon^2 - \delta \right) \lambda^4.
\end{eqnarray}
\end{subequations}
Therefore, thanks to the Biermann-Weierstrass theorem, we can write the formula for $\xi = \xi(s)$ as 
\begin{equation}
\label{xi_s}
    \xi(s) =  \xi_0 + \frac{- \epsilon_{r} \sqrt{f(\xi_0)} \wp'(s) + \frac{1}{2} f'(\xi_0 ) \left[ \wp(s) - \frac{1}{24}f''(\xi_0 )\right] + \frac{1}{24} f(\xi_0 ) f'''(\xi_0 )  }{2 \left[ \wp(s) - \frac{1}{24} f''(\xi_0 ) \right]^2 - \frac{1}{48} f(\xi_0 ) f^{(4)}(\xi_0 ) }.
\end{equation}
Here $\wp$ is understood to be defined by the invariants $g_2$, and $g_3$ given by Eq.\ (\ref{invariants_phys}), i.e., $\wp(z) = \wp(z;g_2,g_3)$, and $f$ is defined in Eqs.\ (\ref{f_general}) and (\ref{fcoeffs}).

The above equation is a general solution to the Eq.\ \eqref{equationf}, valid for all types of allowed trajectories. One can also show that Eq.\ (\ref{xi_s}) gives a correct solution also in the presence of turning points. In this case the sign $\epsilon_r$ in Eq.\ (\ref{xi_s}) has to be interpreted as referring to the initial position $\xi_0$ (in other words, $\epsilon_r$ can be treated as a part of initial data).

\subsection{Solution for \texorpdfstring{$\theta(s)$}{0(s)} and \texorpdfstring{$\varphi(s)$}{f(s)}}

Equation \eqref{theta_mot} can be integrated in a straightforward way. Substituting $\mu = \cos \theta$, we get
\begin{eqnarray*}
    s  & = & \epsilon_\theta \int^\theta _{\theta_0} \frac{1}{\sqrt{ \lambda^{2} - \frac{\lambda_{z}^2}{\sin^2 \tilde{\theta}}}} d\tilde\theta = -\frac{\epsilon_\theta}{\lambda} \int^{\mu}_{\mu_0} \frac{1}{\sqrt{1-\frac{\lambda_z^2}{\lambda^2} - \tilde{\mu}^2 }} d\tilde\mu \\
    & = &  -\frac{\epsilon_\theta}{\lambda} \left( \arcsin{\frac{\mu}{\sqrt{1-\frac{\lambda_z^2}{\lambda^2}}}}  - \arcsin{\frac{\mu_0}{\sqrt{1-\frac{\lambda_z^2}{\lambda^2}}}} \right),
\end{eqnarray*}
where $\mu_0 = \cos \theta_0$ correspond to a reference point. Setting
\[
\beta_0 = \frac{\epsilon_\theta}{\lambda} \arcsin{\frac{\mu_0}{\sqrt{1-\frac{\lambda_z^2}{\lambda^2}}}},
\]
one gets
\begin{equation}\label{Eq_mu}
    \mu(s)= - \epsilon_\theta \sqrt{1-\frac{\lambda_z^2}{\lambda^2}} \sin \left[ \lambda(s-\beta_0) \right].
\end{equation}

Equation \eqref{varphi_mot} can be now integrated as
\begin{eqnarray*}
    \varphi(s) - \varphi_0 &=& \lambda_z  \int^s_0 \frac{1}{\sin^2 \theta(\tilde{s})} d\tilde{s}= \lambda_z  \int^s_0 \frac{1}{1 - \mu(\tilde{s})^2} d\tilde{s} \\
    &=& \arctan\left[ \frac{\lambda_z}{\lambda} \tan [\lambda(s-\beta_0)] \right] + \arctan\left[ \frac{\lambda_z}{\lambda} \tan (\lambda \beta_0) \right],
\end{eqnarray*}
where we have used Eq.\ (\ref{Eq_mu}), and $\varphi_0 = \varphi(0)$. The above expression can be corrected for discontinuities due to the tangent function.  The final formula has the form
\begin{equation}\label{Eq_varphi}
     \varphi(s) =  \varphi_0 + \arctan\left[ \frac{\lambda_z}{\lambda} \tan[\lambda(s-\beta_0)] \right] + \arctan\left[ \frac{\lambda_z}{\lambda} \tan(\lambda \beta_0) \right] + n \pi \sgn \lambda_z,
\end{equation}
where $n$ is an integer part of $\frac{\lambda}{\pi}(s-\beta_0) \pm \frac{1}{2}$. Here the plus corresponds to $s > 0$ and the minus to $s < 0$.

\subsection{Solution for \texorpdfstring{$\tau(s)$}{t(s)} }

A direct integration of Eq.\ (\ref{tau_mot}) yields
\begin{eqnarray*}
    \tau(s) - \tau(0) & = & \varepsilon \int^s_0 \frac{ \xi(\tilde{s})^2}{1-\frac{2}{\xi(\tilde{s})}} d\tilde{s} =  \varepsilon \int^s_0 \left( \xi(\tilde{s})^2 + 2 \xi(\tilde{s}) + 4 +  \frac{8}{\xi(\tilde{s}) -2}  \right)d\tilde{s}.
\end{eqnarray*}
These integrals can be computed by substituting expression (\ref{xi_s}) for $\xi(s)$, however the result may be too complicated and impractical. Instead, we will focus on a simpler case, in which a turning point can be chosen as a reference point.

Let $\xi_1$ denote the radius of the turning point, so that $f(\xi_1) = 0$. Choosing $\xi_1$ as a reference radius corresponding to $s = 0$, one can rewrite Eq.\ (\ref{xi_s}) as
\begin{equation}
\label{xi_s_simp}
    \xi(s) =  \xi_1 + \frac{ \frac{1}{4} f'(\xi_1) }{\wp(s) - \frac{1}{24} f''(\xi_1)},
\end{equation}
irrespectively of the radial direction of motion, i.e., the value of $\epsilon_r$. The coordinate time elapsed during the motion from $ s = 0$ to $s = s_2$ can be written as 
\begin{eqnarray}
     \tau_\ast(s_2;\xi_1) = \varepsilon \int_0^{s_2} && \left\{ \frac{\xi_1^3}{\xi_1 - 2} + \frac{\frac{\xi_1 +1}{2}f'(\xi_1) }{\wp(\tilde{s}) -\wp(y) } - \frac{\frac{2  }{\left( \xi_1 - 2 \right)^2} f'(\xi_1)}{\wp(\tilde{s}) -\wp(z)} + \right. \nonumber \\
    && \left.  \frac{\frac{ 1}{16}f'(\xi_1)^2 }{\left[ \wp(\tilde{s}) -\wp(y) \right]^2}    \right\} d\tilde{s},
    \label{tauast}
\end{eqnarray}
where $\wp(y) = \frac{1}{24} f^{\prime \prime}(\xi_1)$  and  $\wp(z) = \frac{1}{24} f^{\prime \prime}(\xi_1) - \frac{f'(\xi_1)}{4(\xi_1 - 2)}$. One can choose any values $y$ and $z$ satisfying these conditions. Integral (\ref{tauast}) can be computed with the help of the following two integral formulas (\cite{byrd_handbook_1971}, p. 312 and \cite{gradshtein_table_2007}, p. 626):
\begin{eqnarray}
I_1 (x;y) & = & \int \frac{dx}{\wp(x) - \wp(y)} = \frac{1}{\wp'(y)} \left[ 2\zeta(y) x + \ln{\frac{\sigma( x - y)}{\sigma(x + y)}} \right], \label{I_1}\\
I_2 (x;y) & = & \int \frac{dx}{\left(\wp(x) - \wp(y)\right)^2} = - \frac{\wp''(y)}{\wp'^3(y)}\ln{\frac{\sigma\left(x - y\right)}{\sigma\left(x + y\right)}} \nonumber \\
&& -\frac{1}{\wp'^2\left( y\right)} \Bigg\{ \zeta\left(x +y\right) + \zeta\left(x - y\right) + \left[ 2\wp\left( y\right) + \frac{2 \wp''\left( y\right)\zeta\left(y\right)}{\wp'\left(y\right)} \right] x \Bigg\}, \nonumber \\ \label{I_2}
\end{eqnarray}
where $\zeta(x)$ and $\sigma(x)$ denote Weierstrass functions $\zeta(x;g_2,g_3)$ and $\sigma(x;g_2,g_3)$, respectively. Hence we obtain
\begin{eqnarray}
\tau_\ast(s_2,\xi_1) & = & \varepsilon \left\{ \frac{\xi_1^3}{\xi_1 - 2} s_2 + \frac{\xi_1 +1}{2} f'(\xi_1)  \left[ I_1 (s_2;y) - I_1(0;y) \right]  -  \right. \nonumber \\
&& \left.\frac{2 f'(\xi_1) }{\left( \xi_1 - 2 \right)^2} \left[ I_1 (s_2;z) - I_1(0;z) \right] + \frac{f^\prime(\xi_1)^2}{16}  \left[ I_2 (s_2;y) - I_2(0;y) \right] \right\}. \nonumber \\
\label{sastgotowe}
\end{eqnarray}
In general, to obtain a continous expression for $\tau_\ast$, one has to be careful in selecting appropriate branches of the logarithms appearing in Eqs.\ (\ref{I_1}) and (\ref{I_2}). 

As an example, consider a motion of a particle starting from an arbitrary location $\xi_0$ and moving inwards, until it reaches a periapsis with the radius $\xi_1$ (thus $f(\xi_1) = 0$). Next, the particle moves outwards, up to a location with a radius $\xi$. Denote the Mino time needed for the motion from $\xi_0$ to $\xi_1$ by $s_1$ and  the Mino time needed for motion from $\xi_1$ to $\xi$ by $s_2$. Thanks to symmetry, the coordinate time of the entire motion can be written as
\begin{equation}
\label{sgotowe}
\tau(s) = \tau_\ast(s_1;\xi_1) + \tau_\ast(s_2;\xi_1) = \tau_\ast(s_1;\xi_1) + \tau_\ast(s - s_1;\xi_1),
\end{equation}
where  $s = s_1 + s_2$.

The value $s_1$ can be computed from  Eq.\ \eqref{s_integral}. The integral in Eq.\ (\ref{s_integral}) can be expressed in terms of the Legendre elliptic integrals, but the exact form of the result depends on a type of particle trajectory. For an unbound scattered trajectory 
\begin{eqnarray}
    && s(\xi_1) = \epsilon_r \int_{\xi_1}^\infty \frac{d \xi}{\sqrt{f(\xi)}} = \nonumber \\ 
    &&  \frac{\epsilon_r }{\sqrt{y_3 - y_1}} \left[ F \left( \arccos \sqrt{ \frac{y_2 + \frac{1}{12} - \frac{1}{2\xi}}{y_2 - y_1}} , k \right)  - F \left( \arccos \sqrt{\frac{y_2 + \frac{1}{12}}{y_2 - y_1}}, k \right) \right], \nonumber \\
\end{eqnarray}
where $y_1 < y_2 < y_3$ are real zeros of the polynomial $4 y^3 - g_2 y - g_3$, and $k^2 = (y_2 - y_1)/(y_3 - y_1)$. A longer discussion of this result can be found in \cite{CM2022, mach_odrz_accretion_2021}, where it is referred to as the elliptic $X$ function. A comprehensive discussion of the classification of orbits and different expressions for the integral in \eqref{s_integral} can be found in \cite{kostic_analytical_2012,chandrasekhar_mathematical_1983}.

\section{Conclusions}

\tolerance=5000
We have revisited the theory of timelike and null geodesics in the Schwarzschild spacetime. A novel aspect of our study is an application of the Biermann-Weierstrass theorem, providing a single formula describing bound and unbound geodesic orbits. More details on this approach can be found in our recent article \cite{CM2022}. In \cite{CM2022} geodesics are parametrized using the so-called true anomaly---an angle measured within the orbital plane. Here we introduce a parametrization by an analogue of the Mino time, commonly used in the theory of Kerr geodesics. The solutions formulated in this way can serve as limiting expressions for Kerr geodesics in the limit of non spinning black holes.

\section*{Acknowledgments}

This work was partially supported by the Polish National Science Centre Grant No. 2017/26/A/ST2/00530.

\end{document}